\documentclass[sigconf]{acmart}
\AtBeginDocument{%
  \providecommand\BibTeX{{%
    \normalfont B\kern-0.5em{\scshape i\kern-0.25em b}\kern-0.8em\TeX}}}

\setcopyright{rightsretained}
\copyrightyear{2021}
\acmYear{2021}
\acmDOI{}

\acmConference[CHI '21]{CHI '21}
\acmBooktitle{Workshop on Designing Interactions for the Ageing Populations,  May 08--13, 2021, Yokohama, Japan}
\acmPrice{}
\acmISBN{}

\usepackage{adjustbox}

\begin{document}


\title{Computer Anxiety: Supporting the Transition \\ from Desktop to Mobile}



\author{Thiago Donizetti dos Santos}
\affiliation{%
  \institution{Federal University of ABC (UFABC)}
  \city{Santo André, SP}
  \country{Brazil}}
\email{thiagods05@gmail.com}

\author{Vagner Figueredo de Santana}
\affiliation{%
  \institution{IBM Research}
  \city{S\~{a}o Paulo, SP}
  \country{Brazil}
}
 \email{vagsant@br.ibm.com}

\renewcommand{\shortauthors}{Thiago Donizetti dos Santos and Vagner Figueredo de Santana}
\begin{abstract}
Computer Anxiety is a phenomenon studied in multiple contexts and, in the actual COVID-19 scenario, it is gaining more and more importance as it impacts technology adoption and autonomy.
People with Computer Anxiety (PwCA) might feel intimidated, afraid of feeling embarrassed or scared of damaging computers, even before the actual interaction. 
Thus, supporting the detection of Computer Anxiety at scale has the potential to support the technology industry to cope with this challenge.
This position paper presents a user study involving 39 elderly participants in an investigation on the feasibility of using interaction events common to desktop and smartphones to predict different levels of Computer Anxiety. Moreover, it also proposes research directions about the role of smartphones in the context of Computer Anxiety for elderly people as a mean of supporting good first user experiences with technology, meaningful daily use, privacy, and feeling safe even when doing mistakes.
We expect this position paper motivates practitioners, designers, and developers to consider Computer Anxiety as one of the existing barriers when creating mobile applications for elderly people.
\end{abstract}




\maketitle








\section{Introduction}

The use of smartphones in daily activities is increasing in a fast pace. The ownership of smartphones by adults, in US, grew from 35\% to 81\% in the period between 2011 and 2019~\cite{pew2019mobilefactsheet}. Smartphones are generally used to make and receive calls, to access the internet, to text, to access social media and services (e.g., such as food delivery, transport, and mobility). Bearing in mind the ownership of devices, the rate of people owning smartphones drops from 96\% (for the group aging between 18 and 29 years) to 53\% (for the group aging 65+ years) \cite{pew2019mobilefactsheet}. This difference shows that elderly people are increasingly using smartphones, but have not yet adopted this technology in the same way as young adults, which may indicate the existence of aspects impacting the adoption of smartphones by elderly users.

Since a variety of services and content are currently available online, the difficulties faced by elderly people while using smartphones may impact their quality of life. Through the use of mobile applications, one could order food, get a driver using a transport app, travel alone using maps, communicate with family members using instant messaging, learn new things on e-learning platforms or just browse photos and other contents on the social media applications. Such services foment autonomy and help to avoid common stereotypes of dependency and limitations.

One phenomenon that can help in the understanding of issues faced by the elderly people when using new technologies is Computer Anxiety (CA). CA can be defined in terms of affective factors such as intimidation, fear, apprehension, hostility, and worries that one will be embarrassed, will look stupid, or thinks she/he could damage the computer~\cite{heinssen1987assessing}. Although generally related to the use of computers, CA can also impact the use of other electronic devices and previously was also called ``Technophobia'' \cite{balouglu2008multivariate}. CA is related to technology acceptance \cite{Shah2012} and it is generally related to biological changes such as blood pressure, heart rate and electrodynamic responses that occur while a person is using a device \cite{powers1973effects}. CA symptoms can occur during the interaction with the system and even before it, affecting the perceived ease of use and acting as a barrier, impacting the system accessibility as well \cite{santos2018computer}.
Although CA can affect people of all ages, the literature shows that CA is more present in older groups \cite{santos2018computer, chou2009line, fernandez2015age}. Adding this to the relationship between CA and technology acceptance, elderly people may face problems to use mobile devices and, when they use it, CA could make the system difficult to use, make them perform poorly on tasks or fail to achieve their goals while using the device.

In this context, this position paper presents results from a user study involving 39 elderly participants aiming at exploring the feasibility of predicting CA from interaction events common to desktop and smartphones, mapped here as a regression problem. In addition, it also discusses the role of smartphones in supporting people with CA (PwCA) in the process of learning how to use technology, first experience, daily use, and autonomy. Thus, the following research questions were defined to guide the study: \textit{(1) Is it possible to predict different levels of CA using interaction events common to desktop and smartphones?} and \textit{(2) How can smartphones be used to support the adoption of new technologies by PwCA?} The work is organized as follows: the section 2 presents the related work, the section 3 details the user study, the section 4 shows the results, section 5 discusses the role of smartphone for PwCA and section 6 concludes.


\section{Related Work}

\noindent CA is also called Computerphobia, Computer Apprehension, and Technophobia in the literature~\cite{balouglu2008multivariate}.
Rosen et al.~\cite{rosen1987computerphobia} pointed out the following methods and questionnaires to measure CA:

\begin{itemize}
\item \textbf{Computer Anxiety Index} (CAIN) examines avoidance of, caution with, negative attitudes toward, and disinterest in computers~\cite{maurer1984development}.
\item \textbf{Computer Attitude Scale} (CAS) assesses computer liking, confidence, and anxiety through a Likert attitude-measurement format~\cite{loyd1984reliability}. 
\item \textbf{Attitudes Toward Computers Questionnaire} (ATCQ) assesses attitudes towards computer appreciation, usage, and societal impact~\cite{raub1981correlates}.
\item \textbf{Computer Anxiety Rating Scale} (CARS) assesses behavioral, cognitive, and affective components related to technology use~\cite{heinssen1987assessing,rosen1987computerphobia}.
\item \textbf{Mobile Computer Anxiety Scale} (MCAS) assesses anxiety regarding mobile computer using a 38-item Likert scale \cite{wang2007development}.
\end{itemize}

The literature presents multiple factors associated with CA. In sum, PwCA usually have less experience in using computers, have low Computer Self-efficacy (CSE)\footnote{Computer Self-efficacy (CSE) is the belief one has in his/her own abilities to perform a task in the computer~\cite{compeau1995computer}},
take too long to accomplish tasks, perform worse when compared to other users, have negative beliefs about computer/skills, or negative bodily sensations previous/during the interaction with a computer~\cite{santos2018computer}. 
Earlier studies find a strong relationship between age and CA levels, showing evidence that CA is more present in groups with older people and that they have more CA than younger ones \cite{parasuraman1990examination,chou2009line, fernandez2015age}. This can be related to the pace in which technology advances and to the fact that 48\% of older adults report that they usually need someone else to set up a new electronic device or show them how to use it~\cite{pew2017technology}. In this scenario, mobile accessibility has potential to support PwCA in increasing CSE, reducing negative beliefs and worries of using or damaging the device in front of other people.

CA is also present in a few acceptance models. The Technology Acceptance Model (TAM) is one example. It uses the CA as a component that changes the perceived ease of use\footnote{Perceived ease of use is defined as the degree to which a person believes that using a particular system would be free of effort~\cite{abdullah2016investigating}}. So, when considering the adoption of new technology by elderly people, CA should be taken into account, from design to personalization features.

The influence of CA was investigated in contexts such as acceptance of e-learning tools, e-gov, new technologies and health-care systems~\cite{Maki2000,cimperman2016analyzing,Shah2012,jashapara2006understanding,hashim2010paradox,phang2006senior, ivan2016experiencing}. These studies stated that: 
\begin{itemize}
\item PwCA tend to prefer traditional classes instead of e-learning systems and computer-based tests; 
\item PwCA usually perform worse on virtual classes when compared to people without CA; 
\item PwCA have more difficulties in accepting new technologies; 
\item Older PwCA face difficulties using home \textit{telehealth} services and to learn how to use smartphones.
\end{itemize}

Considering support offered to PwCA, the literature presents that instructional/technical support reduce CA in the context of e-learning systems~\cite{gupta2017reducing, mehar2017effect,omogbadegun2019technical}. Finally, smartphones have potential to provide instructional support in a privacy respecting way, supporting the user-technology dialog, reducing worriers associated to trial and error inherent to learning.

\section{Method}

This section details how the user study was run, including its materials, procedure, setup, experiment design, and data analysis planned.
The goal of the study was to collect data related to questionnaires to identify different CA levels and collect detailed interaction data while participants performed tasks on a website in order to detect CA. Moreover, this study also aimed at exploring interaction data common to desktop and smartphone and at understanding the role of smartphone usage for PwCA. 
The types of data captured will be detailed in the Data Analysis section.
Before the main experiment, a pilot was performed in order to assess the user study plan as whole. 
The pilot included 4 elderly participants, recruited the same way the participants of the main experiment (detailed in the next section). The pilot achieved its goals in assessing protocol adopted, duration, and data capture procedure. 
Next, we detail the method followed in the main experiment.


\subsection{Participants}

Elderly people may face difficulties in staying up-to-date with technology and, since they have not used computers since childhood, many of them face CA even when performing a simple task to others age groups~\cite{santos2018computer}. Hence, the target-audience considered in this work is elderly people.

The participants of the experiment were recruited from a list of registered people at the elderly center of the city of 
S\~{a}o Paulo, Brazil, called \textit{Reference Centre for Citizenship of Elderly} (CRECI@). S\~{a}o Paulo is the biggest city in Latin America, with a population of approximately 11.2 million people in the last census (2010) and the current estimate is of 12.2 million people\footnote{https://cidades.ibge.gov.br/brasil/sp/sao-paulo/panorama};
the population of elderly people is approximately 11.9\%\footnote{http://produtos.seade.gov.br/produtos/retratosdesp/view/index.php?\\temaId=1\&indId=4\&locId=3550308}.
Before recruiting participants, a partnership was signed and the proper process for ethics committee was followed at the Federal University of ABC (process \# 2.808.392 and CAEE: 94704418.8.0000.5594), detailing the materials, procedure, questionnaires, data to be collected, and analysis to be performed.


Moreover, only those who had never taken computer classes offered by CRECI@ were invited as potential participants, since results from the literature point that computer classes might reduce CA~\cite{santos2018computer}, which could result in a bias.


\subsection{Materials}

Questionnaires about CA, use of smartphones and computers skills were applied (Appendix \ref{append:questionario}). In order to isolate CA from other comorbidities, questionnaires to assess cognitive abilities and levels of depression were also applied. Only participants with low levels of depression and those who do not present signals of dementia or cognitive deficits had their data considered in the analyses. These two metrics were considered in the exclusion criteria, detailed in the procedure. The questionnaires applied are listed below:

\begin{itemize}
\item Technology use and profile: Has questions about the participant's age, educational level, and frequency of use of computers and smartphones. This questionnaire was applied to give an overview of how participants use technology on a daily basis (Appendix \ref{append:questionario}).

\item Mini Mental: A cognitive screening test used for adults and the elderly to evaluate orientation, memory and attention, naming ability, obedience to verbal and writing commands, free writing of a sentence, and copying a complex drawing (two intersecting polygons). It is currently the most used test for this type of assessment in the world~\cite{MELO2015}. The rationale for using Mini Mental was to identify comorbidity to CA.

\item Geriatric Depression Scale (GDS): GDS is a scale with 30 yes/no questions used for screening depression in elderly people~\cite{ALMEIDA1999,yesavage1982development}. The rationale for using GDS was also to identify comorbidity to CA.

\item Computer Anxiety Rating Scale (CARS): CARS has nineteen questions in a five points Likert scale ranging from strongly disagree to strongly agree. It assesses the behavioral, cognitive and affective components related to technology use~\cite{heinssen1987assessing, rosen1987computerphobia}. The rationale for using CARS is that it is the most referenced questionnaire for screening CA \cite{santos2018computer}. 


\item Computer Self-Efficacy (CSE): CSE is a ten items scale used to assess Computer Self-efficacy~\cite{thatcher2008,compeau1995computer}. The rationale for using CSE was to cross check the results from this study with results from the literature that show that CSE has a strong but inverse relationship with CA.

\item System Usability Scale (SUS): A five points Likert scale questionnaire with ten items. It is often used to assess the perceived usability~\cite{brooke1996sus}. The rationale for using it was to compare perceived usability of the website and the CARS values.
\end{itemize}

In order to capture the interaction data, the participants used a desktop computer including a interaction logger and internet access. The interaction logger used was the open source logger called User Test Logger\footnote{https://github.com/IBM/user-test-logger}. The User Test Logger captures all JavaScript events such as mouse movements, clicks, keys pressed, etc., and generates a raw log file where each line represents an event and information about when, where, and what is related to the event triggered~\cite{santana2018logger}. 


\subsection{Procedure}

The experiment was structured into three steps: pre-test, test, and post-test. The steps are detailed next.

\subsubsection{Pre-test}
\noindent First, screening tests for cognitive deficit, depression, and literacy were applied to identify participants whose scores fall outside the inclusion criteria. The Mini Mental presents a score indicating good cognitive capacity relating the answer points obtained and years of education of the participant as follows: 
\begin{itemize}
    \item No formal education: $\leq 21$ points;
    \item 1 to 5 years of formal education: $\leq 24$ points;
    \item 6 to 11 years of formal education: $\leq 26$ points;
    \item 12+ years of formal education: $\leq 27$ points
\end{itemize}

For GDS screening test, a score of four points or less on the scale indicates low levels of depression. Hence, the exclusion criterion was: $GDS \geq 5$ points. After the tests considered in the exclusion criteria, CARS 
and CSE were applied.


\subsubsection{Test}
\begin{figure}[!ht]
  \centering
  \adjustbox{trim={0.2cm} {0.1cm} {0.2cm} {0.95cm},clip}
  {\includegraphics[width=0.8\columnwidth]{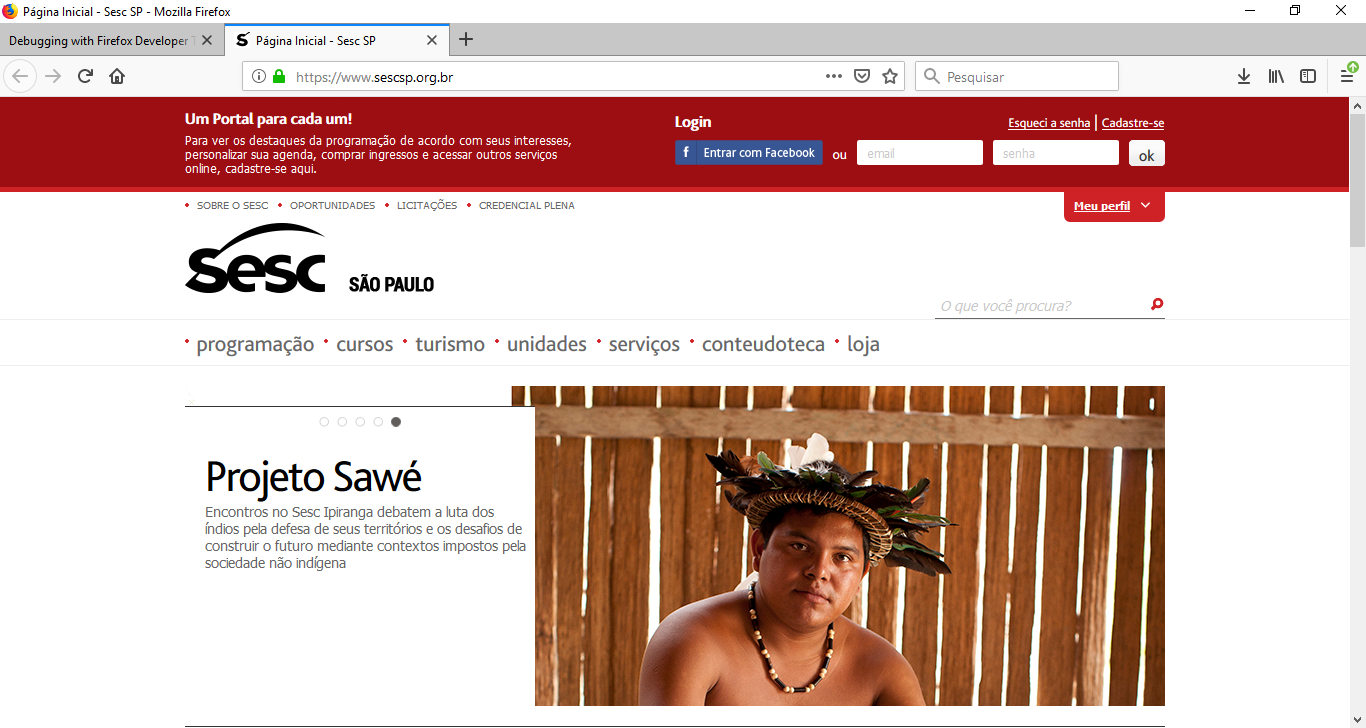}}  
  \caption{Homepage of the SESC's website. 
  }~\label{fig:sesc}
\end{figure}

\noindent The tasks were performed individually on a computer with the User Test Logger installed. 
First, each participant was asked to access SESC homepage (Figure \ref{fig:sesc}). 
The Social Service of Commerce (SESC) is a private entity maintained by the entrepreneurs of the trade in goods, tourism, and services. SESC aims to provide the welfare and quality of life to workers in this sector and their families~\footnote{https://www.sescsp.org.br/}. SESC offers many activities for elderly people such as courses, sports, art exhibition and culture-related events in multiple units in the metropolitan area of S\~{a}o Paulo. The tasks were defined aiming to encourage participants to search online for activities offered by SESC and others services in the city, hoping to help them to see the internet as a tool which they can use as a means of improving their quality of life. The same way they do at CRECI@. In addition, the tasks were structured to be as familiar and as close to real tasks as possible. The tasks read out loud to participants were the following:
\begin{enumerate}
    \item Search for an event, class, or activity he/she might be interested;
    \item Find the address of the unit where the chosen activity/event is offered;
    \item Find the route to the unit.
\end{enumerate}

\vspace{-0.5cm}

There was no maximum time limit for the tasks. Thus, the task duration depended on participants saying whether they finished or gave up on the task. Finally, Thinking-Aloud Protocol~\cite{lewis1982using} was used to understand the rationale of users while performing the tasks. 

\subsubsection{Post-test}
\noindent In order to evaluate the perceived usability and any relationship with CA levels, they were asked to answer the SUS questionnaire.

\subsection{Data analysis}
\noindent According to the exclusion criteria defined, the data from participants who did not score the points required by Mini Mental or scored five or more points on GDS were removed from the data set to be analyzed. 
The resulting data set combined data from the interaction logger, questionnaires, and thinking-aloud protocol. Since the logged data were in raw format, all data captured were processed in order to extract usage metrics. The following metrics were considered having in mind they could also be applied in a mobile interaction setting: time to perform the task, number of clicks and double-clicks, click duration, typing velocity, and total time typing.

Finally, all metrics and the questionnaires scores were combined in a single comma-separated-values (CSV) file used to perform the data analysis. This CSV file is the main data source for the regression analysis performed to predict CARS values based only on interaction data, which could allow its use at scale.  

\section{Results}

\begin{figure}[!ht]
  \centering
    \adjustbox{trim={0.1cm} {0.2cm} {0.1cm} {0.7cm},clip}
    {\includegraphics[width=0.8\columnwidth]{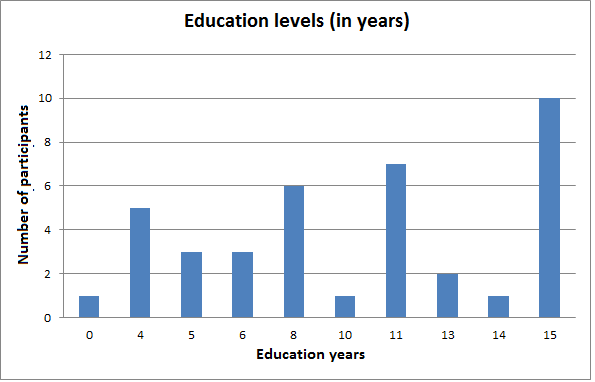}}
  \caption{Distribution of years of formal education.}~\label{fig:escolaridade}
\end{figure}

The experiment included 39 participants, but data from 8 participants were not considered in the analysis due to exclusion criteria. Thus, considering the data from the remaining 31 participants (51.61\% of males, 48.39\% of females). The participants' ages ranged between 62 and 87 years ($\overline{x}$ = 72.84). Regarding the computer usage, 25.81\% of the participants reported that they do not own a computer and do not have frequent access to computers, while 12.90\% do not own a computer, but use it at \textit{lanhouses} or those available in public places; 61.29\% reported owning computers. 
Regarding frequency of use, 61.29\% reported that rarely (less than once a month) use a computer, 16.13\% reported that use it sometimes (more than once a month), 6.45\% reported that usually (more than once a week) use it, and 16.13\% reported that always (everyday) use computers. 
When considering ownership and use of smartphones, 87.10\% of the participants reported that they own smartphones and 48.39\% reported that they always use it. The education level of the participants varies from 0 to 15 years of formal education ($\overline{x}$ = 10.42) (Figure~\ref{fig:escolaridade}).


\vspace{-0.5cm}

The obtained scores for CARS ranged from 20 to 59 ($\overline{x}$ = 42.19). Based on~\cite{cooper2015computer}, the maximum and minimum scores were used to divide the data into 3 groups as follows: $CARS\_range: 59 - 20 = 39$ and $Group\_range: CARS\_range / 3 = 13$.
 \begin{itemize} 
 \item No CA: CARS $<$ 33 (6 participants);
 \item Moderate CA: 33 $\leq$ CARS $<$ 46 (14 participants); 
 \item High CA: CARS $\geq$ 46 (11 participants).
 \end{itemize}

\begin{figure}[!ht]
  \centering
  \includegraphics[width=0.8\columnwidth]{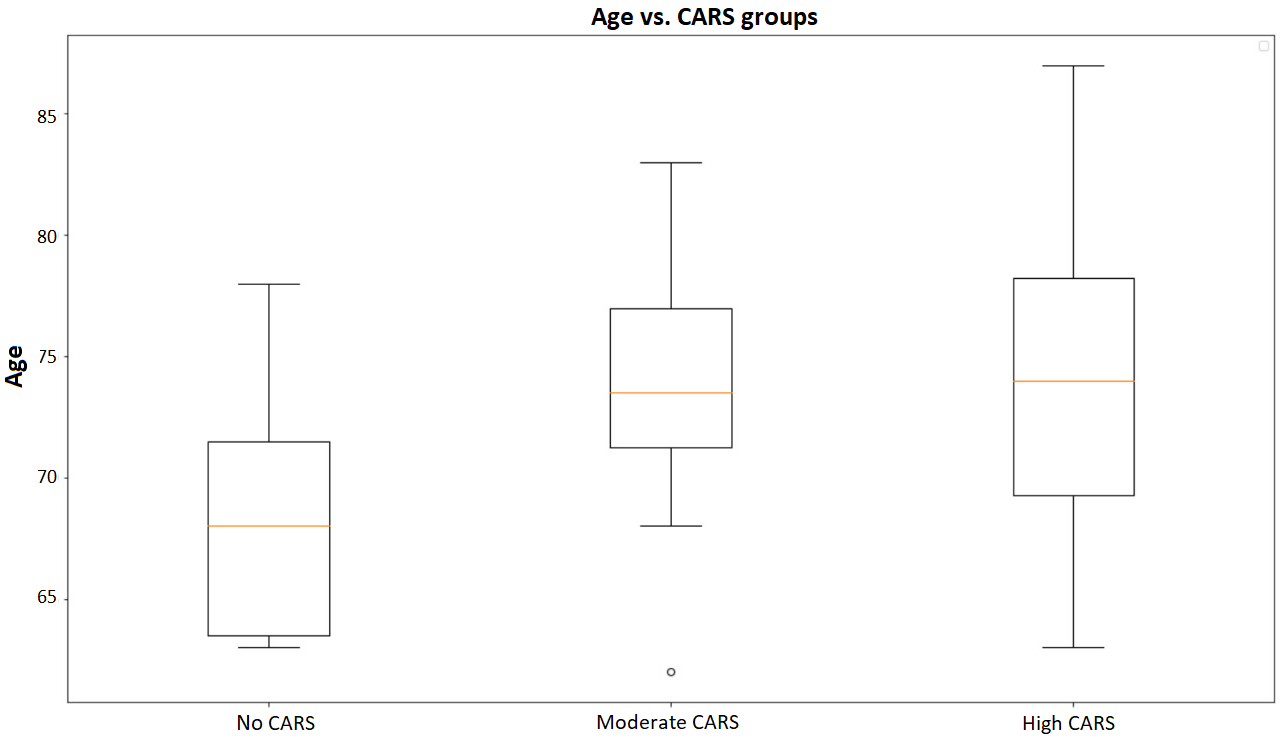}
  \caption{Age distributions for different CA groups.}~\label{fig:ageXCars}
\end{figure}

Considering the three CA groups, Figure~\ref{fig:ageXCars} shows the presence of CA considering the participants' age. It can be seen that participants in the no CA group are among the youngest. The age of these participants ranged from 63 to 78 years old ($\overline{x}$ = 68.43, $\sigma$ = 5.68). The age of the participants in the moderate CA group ranged from 62 to 83 years old ($\overline{x}$ = 73.92, $\sigma$ = 5.60). 
And, for the high CA group, the age of the participants ranged from 63 to 87 ($\overline{x}$ = 74.36, $\sigma$ = 6.77), showing that high levels of CA are present over almost the entire age range covered in the study. Mann-Whitney non-parametric test shows that age was different between no CA and moderate CA groups (p-value = 0.03); no significant difference was found in other pairwise group comparisons.

\vspace{-0.5cm}



\begin{figure}[!ht]
  \centering
	\adjustbox{trim={0.1cm} {0.1cm} {0.1cm} {0.55cm},clip}
	{\includegraphics[width=0.9\columnwidth]{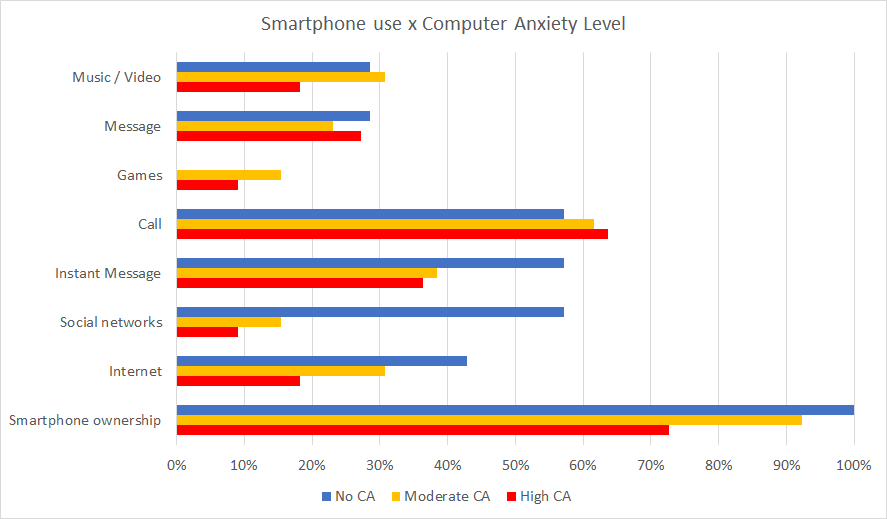}}
  \caption{Smartphone ownership and use by CA groups.}~\label{fig:smartphone-useXCARS}
\end{figure}

Bearing in mind the use of smartphones, Figure~\ref{fig:smartphone-useXCARS} shows smartphone ownership and different uses by different CA groups. It can be seen that high CA group uses smartphones more for calls and less for leisure and other communication activities (e.g., games, music, video, instant messages, and social networks). On the other hand, people in the no CA group use smartphone heavily to access social network, instant messages, and internet.
Analyzing the ownership rate by CA groups, it can be seen that 73\% of the participants with high CA own smartphones, while the ownership is greater in the moderate CA group (92\%) and reaches 100\% for the no CA group. Similarly, considering the frequency of use of computers, people who rarely use computers are the ones with greater CA levels. 82\% of the participants with high CA use it rarely, while it is 61.5\% of the participants of the moderate CA and 28\% of the no CA group.



\vspace{-0.5cm}


Bearing in mind task completion, 23 (74.19\%) participants found an activity and completed the first task. Four participants found an activity that was not of interest to them or was offered by a unit far from their home, but they did not find another one after that. The remaining four gave up without finding an activity. For the Task 2, nine out of 23 (39.13\%) participants found the address of the unit where the selected activity is offered and 14 participants found only the name of the unit.
Regarding the Task 3, six out of nine (66.67\%) participants found the map available on the site, but all of them failed to find the route to the unit. Only two out of nine (22.22\%) participants figured out how to put the starting point address on the map-based UI. 
In sum, task completion dropped from 74.19\% (task 1), to 39.13\% (task 2), and to 0\% (task 3); 22.22\% (6.45\%, considering 31 participants) partially completed the last task. This can be related to task difficulty, fatigue effects, and task dependence. However, after triangulating these results with thinking-aloud data, it was possible to identify that participants faced difficulties with the map-based UI. For instance, participant 11 said: ``It doesn't say where it is. I didn't like SESC.'', participant 35 said: ``Why don't  you have the address on the about the unit page?'' and participant 42 said: ``It will take me a long time to find it (address)''. While using the map, for instance, participant 22 said: ``What should I do here? I have never used it (map) before''; participants are numbered from 1-4 for the pilot and 5-43 for the experiment. 




Table \ref{tab:tasks-mean-time} summarizes the time taken by each group to complete the tasks, showing the mean time and standard deviation by CA group.
It can be seen that the group of participants with high CA had a mean shorter task time than the other groups in some tasks.
This might be related to the fact that PwCA usually gave up more because they feel frustrated, lost, or think that they would not be able to finish the task. This also shows the relationship between high CA and low CSE, as identified in previous studies \cite{santos2018computer} and \cite{santos2019computer}. 
This can be exemplified by the participants quotes as: 
``I think I will have difficulty in this task'', ``This is difficult'', ``I'm lost, I don't know what to do'', and ``I don't know how to find it''. 




\begin{table}[!ht]
  \centering
  \begin{tabular}{l|c|c|c}
     & Task 1 in sec. & Task 2 in sec. & Task 3 in sec. \\ \hline
	Group  & $\overline{x}$ ($\sigma$) & $\overline{x}$ ($\sigma$) & $\overline{x}$ ($\sigma$) \\ \hline
	High CA & 411.18 (230.72) & 599.67 (220.30) & 398.00 (262.19) \\
	Mod. CA  & 647.90 (666.46) & 450.13 (336.52) & 560.50 (30.50) \\
	No CA   & 525.14 (331.39) & 587.00 (390.73) & 399.67 (375.05) \\
 \end{tabular}
 \caption{Average task time by group and standard deviations.}~\label{tab:tasks-mean-time}
\end{table}

\begin{figure}[!ht]
  \centering
  \includegraphics[width=1\columnwidth]{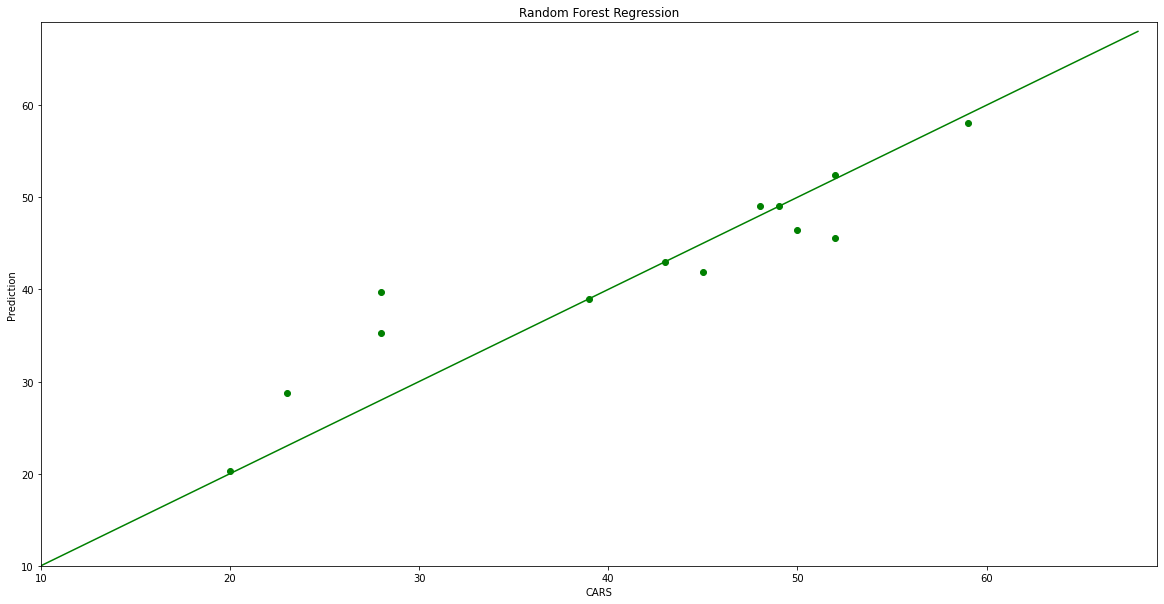}
  \caption{Regression test results of CARS scores prediction.}~\label{fig:cars-regression}
\end{figure}

Although the interaction data was collected during the use of a desktop computer, in this study we explore metrics which can be captured in a mobile setting as well, namely: task time, numbers of clicks and double clicks, mean click duration (interval between pressing and releasing), typing velocity and total time typing. All metrics were normalized for the regression analysis. Prior to fitting the regression model, a random oversampling algorithm was applied\footnote{https://imbalanced-learn.org/stable/over\_sampling.html} addressing the minority values and a train-test split of 80\% / 20\% was applied. Figure \ref{fig:cars-regression}
shows Random Forest regression predictions for CA values (y-axis) vs. CARS values in the test set (x-axis). The obtained regressor has a mean squared error (MSE) of 22.21 and $R^2$ = 0.84. The high MSE value is due to errors related to predictions for lower CA scores. This pessimist prediction would show that users need more support than they would actually need, so such regressor might be useful for indicating when support for PwCA could be applied.

\section{Discussion}
Although there are few studies reporting no significant relationship between age and CA~\cite{parayitam2010computer, korobili2010computer, herrmann2007users}, there are also evidences that older people manifest more CA than younger ones~\cite{parasuraman1990examination, chou2009line, fernandez2015age, russell1997teachers, solberg1998variables}. Results 
suggest that younger participants were in the no CA group, while the older were in the moderate CA group. For high CA group, results suggest that there were participants with high CA almost in the whole age range considered, but it still shows a greater concentration among the older ones (median = 74 years, $\overline{x}$ = 74.2, $sigma$ = 7.12). These results are similar to other findings in the literature, indicating that the CA is more present in the older groups. Moreover, unlike \cite{powers1973effects}, the results suggest that previous experience may impact CA levels, since people who rarely use computers and smartphones were the ones with greater CA levels.

The difficulty in adopting new technologies is suggested by results in Figure \ref{fig:smartphone-useXCARS}. Although there is a high rate of PwCA owning smartphones, the most frequent reported use is making calls. Thus, the participants use the smartphone, but they use the same way they used to do with the old phones: making calls. Moreover, the no CA group presented the same rate (57\%) when using smartphones to make calls, using instant message and social networks apps. In contrast, moderate and high CA groups use more to make calls than to any of the other functions analyzed. Also they use more to make calls then the no CA group and presented a lower rate of ownership of smartphones than the no CA group. The increasing rate of smartphone ownership and the decreasing rate of use of different smartphone functions show possible impacts of CA on the behavior of the elderly regarding technology.
The second most frequent use is instant message apps, even though there is also a difference between groups for this use. The use of this application could be related to the presence of functions such as the ability to make calls using the app or using voice messages. They reported to be used to make calls and that the use of voice messages is easier for them, since they generally have difficulties using the keyboard of the smartphones to write and may have difficulty in reading due to the size of the screen and letters. These findings suggest that high levels of CA may affect people of all ages, but it is more present among older ones due to factors such as lack of practice or lack of knowledge about recent features that could improve UX.

The results regarding task completion and time taken to complete tasks show how CA levels might affect task performance. 
Besides the low task completion rate for the high CA group, 
this group took less time trying to complete the Task 1, showing that PwCA usually gave up more. The implications for HCI researchers in this aspect are related to the design of shorter and simpler tasks and to improve user experience, accessibility and usability, since high CA levels impact negatively the perceived easy of use and CSE, as they may feel frustrated or lost when facing problems to achieve their goals during the interaction.

CA is also related to specific situations since it tends to arise or be stronger during the first use of a device \cite{ivan2016experiencing}. Thus, in addition to promoting the contact of the elderly with new technologies, it is important to promote a good first experience. It means an experience free of effort, that helps the user to feel safe and unafraid of making mistakes. A bad experience, during which the user feels lost or makes mistakes, can reinforce the fears of PwCA. Consequently, this can increase their CA levels, making them believe they are unable to use it and prevent them from trying again.



The results of the Random Forest regression show that CA influences on how users interact with the system. Although the data belongs to desktop computer interaction domain, the result suggests that such approach should be explored in the mobile settings as well, given that the interaction events selected are common to desktop and smartphones. The prediction of higher CARS values could trigger personalization features and additional support, for instance. Smartphones have a myriad of sensors that could be used for such personalization features and here we advocate the use of the following metrics as a starting point: task time, numbers of clicks and double clicks, mean click duration, typing velocity and total time typing. 

This paper defends the idea that the use of smartphones by the elderly can bring benefits such as autonomy, access to content and services and communication. However, CA may create barriers which prevent elderly from enjoying these benefits. Therefore, we argue that further studies are needed regarding the influences of levels of CA in acceptance of smartphones and apps by the elderly people. On his work about the development of a mobile computer
anxiety scale (MCAS), \cite{wang2007development} argues that MCAS is related to three distinct components: (1) traditional CA construct; (2) Internet anxiety construct and (3) special factors making up the mobile anxiety construct (e.g., equipment limitation). The limitations of mobile equipment listed by \cite{wang2007development} and that elderly people report as being problematic for them are: small screens and small multi-function key pads; lower display resolution; unfriendly user-interfaces; and graphical limitations. Thus, these limitations should be considered when developing new technologies for elderly people as well.
Furthermore, the importance of the first experience is found in the literature and reported in the interviews conducted in this study. They reported having purchased or been presented with a new smartphone and feeling lost or afraid to use it. So, it is important that the device has a simple, accessible and usable interface. Another common factor reported, is the fear of making mistakes, looking stupid or breaking the device. In this sense, it is important that the system provides a safe environment, which asks for confirmation for important (dangerous) actions. And, in case the user makes a mistake, the system must provide ways to recover from the error, returning to the previous state without difficulty.

In Brazil, families often have a computer to be shared by family members. It can make elderly people afraid of breaking what belongs to the family or afraid of losing some important data if they do something wrong. The smartphone, on the other hand, is seen as an personal device. According to the participants' reports, this brings greater freedom to learn how to use through trial and error. Besides that, as it is mobile, it has the advantage that elderly people can avoid to use it in front of other people. This can help dealing with the fear of not knowing how to use it or making mistakes in front of younger people. 

Finally, we believe a research agenda about the role of smartphones for elderly people in the context of CA should address the following research questions:
(1) How to detect CA during the use of mobile phones at scale? (2) How to provide (first) good user experiences for this population? (3) How to create a secure environment for PwCA to recover from mistakes? (4) How to combine technology use with learning in order to increase CSE? (5) How CA on the desktop relates to CA on smartphones? 


\section{Conclusion}
This position paper discussed how elderly people use smartphones in a specific region of São Paulo, Brazil, and shows how CA is present in this sample of the population. 
Our findings suggest that higher CA levels are prevalent on higher age and CA impacts how users interact with technologies. In addition, results indicate that the behavior of elderly users when performing tasks can be negatively impacted not only because of age-related factors, but also by the CA levels.
The results indicating the preference of some applications over others by elderly people indicate the need for further studies on why some technologies still present barriers for PwCA. Moreover, the shorter task time obtained by the high CA group and the fact that they usually gave up when feel lost shows the importance of shorter and simpler tasks. The differences between groups regarding ownership of smartphones show that CA may impact on the technology adoption. 
 In addition, the results showing the preference for known functions
can be important to designers and developers to consider when developing new systems, since the inclusion of functions considered easy to use may increase the system adoption and improve user experience by reducing frustration. 

Finally, tackling technology adoption by elderly people may improve their quality of life, since the use of smartphones and the wide variety of applications and services may help they achieve independence, make their lives more comfortable, promoting their better participation in the community and allowing access to the most diverse online content. The presented user test shows that metrics such as task time, number of clicks, click duration, typing speed and total time typing can support the prediction of different CARS scores ($R^2$=0.84). In the current context of the COVID-19 pandemics, promoting autonomy, communication and leisure activities became core goals for any technology and here we emphasize that by understanding CA and considering it in design and development phases mobile apps have the potential to change the live of PwCA.


\begin{acks}
    We thank the CRECI@ for all the support.
    This study was financed in part by the Coordenação de Aperfeiçoamento de Pessoal de Nível Superior - Brasil (CAPES) - Finance Code 001.
\end{acks}

\bibliographystyle{ACM-Reference-Format}
\bibliography{sample-base, computer_anxiety_2020}

\appendix

\section{Technology use and profile}
\label{append:questionario}

\begin{enumerate}
\item Do you have a computer available at home? If not, do you use a computer somewhere else (e.g., at work or in the lanhouse)?
\item How often do you use a computer?\\
A: [  ] Rarely \hspace{0.2cm} [  ] Sometimes \hspace{0.2cm} [ ] Usually \hspace{0.2cm} [  ] Always
\item What do you usually do on computer?
\item Do you own a smartphone?
\item How often do you use smartphones?\\
A: [  ] Rarely \hspace{0.2cm} [  ] Sometimes \hspace{0.2cm} [ ] Usually \hspace{0.2cm} [  ] Always
\item What do you usually do on smartphone?\\
Internet: [  ] Yes \hspace{0.2cm} [  ] No\\
Social networks: [  ] Yes \hspace{0.2cm} [  ] No\\
Instant Message: [  ] Yes \hspace{0.2cm} [  ] No\\
Call: [  ] Yes \hspace{0.2cm} [  ] No\\
Games: [  ] Yes \hspace{0.2cm} [  ] No\\
Message: [  ] Yes \hspace{0.2cm} [  ] No\\
Music / Video: [  ] Yes \hspace{0.2cm} [  ] No\\
\end{enumerate}




\end{document}